\documentclass[a4paper,10pt]{article}

\usepackage{geometry}
\usepackage{amssymb}
\usepackage{algorithm}
\usepackage{algpseudocode}
\usepackage{amsmath}
\usepackage{url}
\usepackage{subfigure}
\usepackage{graphicx}
\graphicspath{{res/}}

\usepackage[numbers]{natbib}
\renewcommand{\cite}{\citep}

\newcommand{\fref}[1]{Figure \ref{#1}}
\newcommand{\aref}[1]{Algorithm \ref{#1}}
\newcommand{\tref}[1]{Table \ref{#1}}

\begin{document}

\title{Fast Multi-Scale Community Detection based on Local Criteria within a Multi-Threaded Algorithm}

\author{Erwan Le Martelot\\
\small Department of Computing\\
\small Imperial College London\\
\small  London SW7 2AZ, United Kingdom\\
\small \texttt{e.le-martelot@imperial.ac.uk}\\
\and
Chris Hankin\\
\small Department of Computing\\
\small Imperial College London\\
\small  London SW7 2AZ, United Kingdom\\
\small \texttt{c.hankin@imperial.ac.uk}
}

\maketitle

%%%
% NOTES
%
% In the conclusion, elaborate more on complexity result.
%%%%

\begin{abstract}
Many systems can be described using graphs, or networks. Detecting communities in these networks can provide information about the underlying structure and functioning of the original systems. Yet this detection is a complex task and a large amount of work was dedicated to it in the past decade. One important feature is that communities can be found at several scales, or levels of resolution, indicating several levels of organisations. Therefore solutions to the community structure may not be unique. Also networks tend to be large and hence require efficient processing. In this work, we present a new algorithm for the fast detection of communities across scales using a local criterion. We exploit the local aspect of the criterion to enable parallel computation and improve the algorithm's efficiency further. The algorithm is tested against large generated multi-scale networks and experiments demonstrate its efficiency and accuracy.
\end{abstract}

\paragraph{Keywords:} 
community detection, local criterion, multi-scale, fast computation, large networks, parallel computation, multi-threading

%%%%%%%%%%%%%%%%%%%%%%%%%%%%%%%%%%%%%%%%%%%%%%%%%%%%%%%%%%%%

\section{Introduction}
\label{sec:introduction}

Social interactions, Internet, telephone networks, power grids, transportation networks, protein interactions, all have in common that they can be represented and studied as graphs, or networks \cite{Newman2010}. Network science grew to become a wide-reaching field where advances impact many others fields.
In the past decade the field of community detection attracted a lot of interest considering community structures as important features of real-world networks \cite{Fortunato2010}. Commonly, community detection refers to finding groups of nodes more densely connected internally than externally. As opposed to clustering methods which commonly involve a given number of clusters, communities are usually unknown, can be of unequal size and density, and can be hierarchical \cite{Fortunato2010,Leskovec2008}. Finding communities can provide information about the underlying structure of a network and its functioning. It can also be used as a more compact representation of the network, for instance for visualisations.

Techniques to uncover communities may consider the network as a whole (global perspective) or may explore smaller areas progressively through their neighbourhoods (local perspective). Usually global techniques run faster but impose crisp boundaries while local techniques are slower but allow overlapping communities.
Also scale parameters can be used to bias the detection towards clusters of various sizes. Community detection can therefore be approached in several ways. This resulted in the creation of various methods to address the problem \cite{Fortunato2010,Leskovec2010}.
In general, community detection methods use a criterion to rank communities and an optimisation algorithm to process the data. These criteria consider either a global or a local perspective. The algorithms often rely on heuristics in order to process the data in a reasonable amount of time. Indeed the division into communities of a network is an NP-hard task \cite{Fortunato2010} and datasets in real-world problems are often large. Therefore a significant emphasis must be put on producing algorithms with a low complexity.
Also networks often have several levels of organisation \cite{Simon1962}, leading to different relevant communities at various scales (or resolutions). Accurate community detection in a network therefore implies uncovering communities at identified scales of relevance.

Recently \cite{LeMartelot2013a} addressed this issue and introduced a method for the efficient detection of communities across scales on large networks. This method was implemented by two algorithms, respectively designed for global and local criteria. Both algorithms can handle large graphs, with only the local criteria one enabling overlapping communities. Yet the local criteria algorithm has a greater complexity, polynomial, compared to the global criteria algorithm which has a linear complexity. Therefore the performance of the local criteria algorithm is significantly inferior to the performance of the global criteria one and its scalability is reduced. While enabling overlapping communities increases the complexity of the task it would still be desirable to reach a scalability comparable to the one of the global criteria algorithm.

To address this we focus in this work on the local criteria approach and present an algorithm implementing the method from \cite{LeMartelot2013a} with an improved efficiency. The algorithm also exploits features of local criteria to enable multi-threading at its core.

The following section reviews the relevant contributions found in the literature. Then our new algorithm is presented. It is followed by experiments performed on large networks and conclusions.

%%%%%%%%%%%%%%%%%%%%%%%%%%%%%%%%%%%%%%%%%%%%%%%%%%%%%%%%%%%%

\section{Background}

In the recent years several multi-scale criteria and associated methods to uncover communities were introduced \cite{Reichardt2006,Arenas2008,LeMartelot2012,Ronhovde2010,Lancichinetti2009b,Huang2011}. Based on some of these criteria, a new method for the fast detection of communities across scales was introduced in \cite{LeMartelot2013a}. Given an ordered sequence of scale parameters, this method considers that the outcome of the algorithm for a specific parameter value is valuable information that can be exploited for further parameter values. More specifically the result for parameter value $p$ is used to uncover the result for the following parameter $p+\delta p$. The method therefore exploits the input data and the information computed as the algorithm runs.

Initially the method was derived into two algorithms, one for global criteria and one for local criteria. However in a local criteria approach communities are grown independently which makes it naturally suited to parallel computation. In this work we will only consider the local criteria algorithm. Another asset of the local criteria approach is that due to the independence of the growth process between communities, the resulting communities can share nodes and thus be overlapping. This is a feature that the global criteria approach does not provide.

The method is based on an aggregation process that builds larger and larger communities as parameters are given in order of increasing scale. The input parameter list must be such that $\forall (i,j) \in \mathbb{N}^2: i < j \Rightarrow scale(p_i) < scale(p_j)$ where $scale(p)$ represents the coarseness level of the scale parameter value $p$. The larger the value, the coarser the scale. For each parameter $p_j$ following $p_i$ the algorithm will start its computation based on the outcome for $p_i$ instead of starting from scratch.
To deal with small variations as well as larger variations between successive sets of communities, the method uses two phases. One phase performs subtle changes at the node level. The second phase performs coarser operations at the community level. These phases alternate until no further refinement is possible for a given scale parameter. Then the method uses the current outcome as a starting point for the next scale.
The first phase of subtle changes is performed using a growth function that expands communities until no further improvement of the criterion can be made. The larger change phase merges communities that overlap significantly, thus reducing the amount of communities while maintaining their integrity.

Initially the method was implemented for two local criteria: the criterion from \citeauthor{Lancichinetti2009b} \cite{Lancichinetti2009b} and the the criterion from \citeauthor{Huang2011} \cite{Huang2011}. However experiments showed that the criterion from \cite{Lancichinetti2009b} was more efficient and faster to optimise. We therefore chose here to reuse this criteria.
In \cite{Lancichinetti2009b} the authors introduced the fitness of a community $c$ as
\begin{equation}
	f_c = \frac{k^c_{in}}{(k^c_{in}+k^c_{out})^{\alpha}}
	\label{eq:lfk_com_fit}
\end{equation}
and then test whether a node $i$ should join a community $c$ by computing the fitness of $i$ with respect to $c$ as
\begin{equation}
	f^i_c = f_{c+i} - f_{c-i}
	\label{eq:lfk_node_fit}
\end{equation}
The parameter $\alpha$ sets the scale of the method. Large values of $\alpha$ lead to small communities while small values lead to large ones. We will hereafter call this fitness function the LFK criterion.

The criterion is used in the growth phase. The idea for growing communities is to start from an initial node called seed or from an existing community and then grow the community by successively adding neighbour nodes that improve the criterion value until no node can be added.
%We will use a growth phase optimised for LFK derived from the one presented in \cite{LeMartelot2013a}.
Candidate nodes for joining the community are considered in order using a max priority queue with ranking factor $\frac{2.d_{in}}{(d_{in}+d_{out})^{\alpha}}$ to rank nodes, where $d_{in}$ is the sum of edge weights from a node to a community and $d_{out}$ the remaining edge weights of the node. Once all possible nodes have been added, the algorithm checks whether the member nodes of the community still contribute to the criterion improvement. If they no longer do, they are removed. 
The growth algorithm is given in \aref{alg:lfk_growth}.
\begin{algorithm*}[hbt]
	\caption{Fast method from \cite{LeMartelot2013a} to grow a community $c$ using the criterion from \cite{Lancichinetti2009b}.}
	\label{alg:lfk_growth}
	\footnotesize{
	\begin{algorithmic}[1]
		\State Create neighbour nodes max priority queue using factor $\frac{2.d_{in}}{(d_{in}+d_{out})^{\alpha}}$
		\While{priority queue is not empty}
			\State Pick first node $n$
			\If{$n$ improves $Q_c$}
				\State Add $n$ to $c$
				\State Update or add in priority queue neighbours of $n$ not in $c$ 
			\EndIf
		\EndWhile
		\If{a node has been added}
			\While{Number of iterations $< k$}
				\ForAll{nodes $n$ in $c$}
					\State Recompute $Q_{c \backslash n}$
					\If{$Q_{c \backslash n} > Q_c$}
						\State $n$ is removed from $c$
					\EndIf
				\EndFor
				\State Exit $while$ loop if no node could be removed
			\EndWhile
		\EndIf
	\end{algorithmic}
	}
\end{algorithm*}

Regarding the merging phase, local criteria are not suitable. They are designed to consider the addition or removal of nodes to a community in order to perform a growth process. They are not designed to assess larger operations such as the merging of several communities. Therefore the second phase merges communities if they overlap significantly. As communities grow independently from one another in the first phase, some may overlap. The overlap ratio for merging is controlled by a threshold $\eta$. Two communities $C_1$ and $C_2$ are merged if $max(\frac{|C_1 \cap C_2|}{|C_2|}, \frac{|C_1 \cap C_2|}{|C_1|}) \ge \eta$. ($|C|$ refers to the cardinality of $C$.) By default  $\eta = 0.5$ so a community merges into another one if at least half of the nodes belong also to the other one.

Parallel computing has also recently been used for community detection. In \cite{Riedy2012b} the authors presented a crisp community detection algorithm for the optimisation of modularity or conductance. Their implementation enables fast computation but relies on specific hardware with massive parallelism and is therefore not easily portable.
Another approach using parallel computation was presented in \cite{Soman2011}. The authors present an algorithm based on the label propagation algorithm \cite{Raghavan2007} using GPGPU. Their experiments demonstrate the speed efficiency of their approach. 
These two approach provide good insights into the usage of parallelism in community detection methods, yet they are more focussed on parallelism than usability and accuracy. Also both approaches ignore the multi-scale aspect of communities in real-world data.

In this work we design a new algorithm following the steps of the method from \cite{LeMartelot2013a}. However we add a focus on parallelism to speed up the algorithm while making it usable by any user. The algorithm is designed to exploit the parallelism offered by the multi-core architecture present in most recent computers.

%%%%%%%%%%%%%%%%%%%%%%%%%%%%%%%%%%%%%%%%%%%%%%%%%%%%%%%%%%%%

\section{Algorithm}

Local approaches have the advantage of only working with local information. Each area of interest in a network can thus potentially be explored independently from others. This distribution of tasks suits a parallel computation approach. Therefore we present a new algorithm implementing the method from \cite{LeMartelot2013a} and making extensive use of parallel computation. The pseudo-code is given in \aref{alg:mscd_local_lfk2}.
%\footnote{A C++ implementation is available for download from \url{www.elemartelot.org}. It has been integrated to the framework presented in \cite{LeMartelot2013a}.}

\begin{algorithm*}[!hbt]
	\caption{Parallel multi-scale community detection algorithm for local criteria.}
	\label{alg:mscd_local_lfk2}
	\footnotesize{
	\begin{algorithmic}[1]
	
\If{a set of initial communities is given in input}
	\State Set it as the current set of communities
\Else
	\State Initialise all nodes with a least 2 connections as potential seeds: $seedset =$ set of all seeds
	\While{$seedset$ is not empty}
		\State Initialise new community $c$ with a seed $n$
		\State Remove from $seedset$ the seed $n$ and all its neighbours
		\If{second seed rule applies}
			\State Remove from $seedset$ the neighbours of neighbours of $n$
		\EndIf
	\EndWhile
\EndIf
	
\ForAll{scale parameters $p$}
	\While{changes can be made}
		\State Reinitialise the list of node membership sets $memsets$
		\State Split the set of communities into $t$ distinct subsets and launch $t$ threads
		\ForAll{communities $c$ in the thread subset} \Comment Running in each thread
			\State Grow $c$ according to the criterion tuned by $p$
			\If{$c$ changed}
				\State Add $c$ to the set $C_C$ of communities to check for merging
			\EndIf
		\EndFor
		
		\State Split the set of communities to check $C_C$ into $t$ distinct subsets and launch $t$ threads
		\ForAll{communities to check $c$ in the thread subset} \Comment Running in each thread
			\State Initialise for each community a counter $count$ of nodes shared with $c$
			\ForAll{nodes $n$ in $c$}
				\ForAll{communities $c_n$ in $memsets[n]$}
					\State Increment $count[c_n]$
					\If{$count[c_n]$ reaches the merging threshold}
						\State Add the pair of communities $(c,c_n)$ to the set of communities to merge $C_M$
						\State Exit loop for current community $c$
					\EndIf
				\EndFor
			\EndFor
		\EndFor
		
		\While{the set of communities to merge $C_M$ is not empty}
			\State Split in $t$ distinct subsets the pairs that have no community overlap
			\State Remove the pairs from $C_M$
			\State Launch $t$ threads
			\ForAll{pair of communities ($c_1,c_2$) in the thread subset} \Comment Running in each thread
				\State Merge communities into $c_1$
				\State All references to $c_2$ on the merge set are renamed $c_1$
			\EndFor
		\EndWhile
	\EndWhile
	\State Store community set and $Q$ for $p$
\EndFor
\State \Return{Community sets and associated $Qs$}
	\end{algorithmic}
	}
\end{algorithm*}

The algorithm is initialised with a set of nodes called seeds that will form the initial communities. Note that precomputed communities can also be given instead. Seeds are selected randomly from a candidate set, removed from it and added to the seed set. All the neighbours of this seed are then removed from the set of remaining seed candidates. This prevents starting different communities from neighbour nodes which would very likely result in similar communities and hence waste computing resources.
A second rule can consider discarding also the neighbours of neighbours and thus guarantees a minimum of two intermediate nodes between two seeds. As each seed will be a community to process the number of seeds chosen initially impacts the runtime of the algorithm. Therefore reducing the number of seeds is important. However it may also reduce the accuracy of the algorithm.

Once communities have been initialised the algorithm begins its loop through all scale parameters. For each scale, while changes can be made the algorithm keeps analysing the current scale. The implementation from \cite{LeMartelot2013a} follows two phases. In the first one communities are grown. In the second one significantly overlapping communities are merged. We keep these two phases here with some modifications. First communities are grown in parallel. When a community is modified it is then added to a list of communities to check for merging. The second phase consists of the checking and merging steps. All the communities on this check list are processed in parallel to find whether they overlap beyond a merging threshold. When two communities overlap enough the pair is added to a merge list. Finally the merge list is processed. All pairs that have no community in common are merged in parallel. Then references are updated in the remaining communities to merge (e.g. if $c_2$ merged in to $c_1$, references to $c_2$ are renamed $c_1$) and the parallel merging process is repeated until all pairs of communities have been merged.

Regarding the growth function, in order to prevent the growth of communities already overlapping significantly with others we added a test at the beginning of the growth function from \aref{alg:lfk_growth}. The test checks the amount of shared nodes with each overlapping communities. If an overlap reaches the merging threshold then the growth function returns such that \aref{alg:mscd_local_lfk2} on lines 19 to 21 adds the community to the list of communities to check for merging. The community still requires further checking as after all communities have been grown, their structure may have changed and the merging may no longer be required.

The community memberships are maintained and updated in a list of membership sets. Each node has its own community membership set. These sets are updated each time a node is added to a community or removed from one. As several growth functions run simultaneously the memberships may be requested concurrently for reading and writing. Therefore we implemented these membership sets as atomic sets using the readers-writers problem solution with priority to writers (second R/W solution) from \cite{Courtois1971}. The modified growth function we use here is given in \aref{alg:lfk_growth_2}.

\begin{algorithm*}[hbt]
	\caption{Modified growth function.}
	\label{alg:lfk_growth_2}
	\footnotesize{
	\begin{algorithmic}[1]
		\State Initialise shared node counter for communities to 0
		\For{all nodes $n$ in the community}
			\State Get community membership of $n$ from membership atomic set $memset[n]$
			\For{all communities $c$ of $n$}
				\If{$c$ is not the current community being grown}
					\State Increment the node counter for $c$
					\If{the counter for $c$ reaches the merge threashold}
						\State Return true (for \aref{alg:mscd_local_lfk2} on lines 19-21)
					\EndIf
				\EndIf
			\EndFor
		\EndFor
		
		\State Create neighbour nodes max priority queue using factor $\frac{2.d_{in}}{(d_{in}+d_{out})^{\alpha}}$
		\While{priority queue is not empty}
			\State Pick first node $n$
			\If{$n$ improves $Q_c$}
				\State Add $n$ to $c$
				\State Add $c$ to $memset[n]$
				\State Update or add in priority queue neighbours of $n$ not in $c$ 
			\EndIf
		\EndWhile
		\If{a node has been added}
			\While{Number of iterations $< k$}
				\ForAll{nodes $n$ in $c$}
					\State Recompute $Q_{c \backslash n}$
					\If{$Q_{c \backslash n} > Q_c$}
						\State $n$ is removed from $c$
						\State Remove $c$ from $memset[n]$
					\EndIf
				\EndFor
				\State Exit $while$ loop if no node could be removed
			\EndWhile
		\EndIf
		
	\end{algorithmic}
	}
\end{algorithm*}

In the growth process, we reused the local criterion (LFK) from \cite{Lancichinetti2009b} based on the results from \cite{LeMartelot2013a}. Any other local criterion could be used though.

\paragraph{Complexity Analysis:}

The seeds initialisation run in $\mathcal{O}(n \cdot d)$ where $n$ is the number of nodes and $d$ the average degree of a node. Using the second seed rule it runs in $\mathcal{O}(n \cdot d^2)$. Then the algorithm runs through all scale parameters $p$. This number of parameters being small in front of $n$, $p$ does not affect the overall complexity. For each parameter the algorithm loops as long as changes can be made, which enables the alternation of the growth and the merging phases. In practice this loop is repeated only a few times. The reinitialisation of the membership list is done in $\mathcal{O}(n)$ by scanning through all nodes in all communities. Preparing the additional data needed by the threads is done in constant time.

The complexity of the growth process is difficult to evaluate. The first phase testing the overlapping with other communities runs in $\mathcal{O}(n_k \cdot n_c \cdot r)$ where $n_k$ is the average community size for a given scale, $n_c$ is the number of communities and $r$ represents the ratio of overlapping communities. If this ratio is low, it runs in $\mathcal{O}(n_k)$.
The creation of the priority queue is done in $\mathcal{O}(n_k \cdot d)$. Note that the direct access to the set of neighbour nodes of a community requires the maintenance of a neighbours set structure for each community. This maintenance requires a few additional operations during the growth process.
Then for all nodes $n_k$ in the queue, the LFK criterion is calculated iterating through the $d$ edges (on average) of each node. If a node is added the queue is amended in up to $\mathcal{O}(d^2)$ as each neighbour of the added node may be added to the queue and iterating through its edges is required to compute the ranking factor. Therefore this part runs in up to $\mathcal{O}(n_k \cdot d^2)$. As in practice not all nodes are added the complexity is lower.
The final set of loops checking whether a node should still belong to the community is performed in practice only a few times. The inner loop iterates through the $n_k$ nodes and computes the criterion value of the node in $d$ steps. If the node is removed, up to $\mathcal{O}(d^2)$ operations are needed to update the set of neighbours. Therefore this part also runs in $\mathcal{O}(n_k \cdot d^2)$ but again, in practice, not all neighbours are removed.
The growth process therefore runs between $\mathcal{O}(n_k \cdot d)$ and $\mathcal{O}(n_k \cdot d^2)$ for each community. A quick sort at the end of the function is used to keep the community nodes sorted. This operation has a complexity of $\mathcal{O}(n_k \cdot log(n_k))$. As this remains in the scale of the previous complexity ranges, it will be ignored.

The complexity of the checking and merging parts may vary. Checking if two communities overlap significantly is done in linear time, as well as merging them. The theoretical worst case is when all communities are checked against all the other communities, in which case the complexity reaches $\mathcal{O}(n_c^2 \cdot n_k)$. In practice communities are only checked against their neighbour communities, bringing the complexity to $\mathcal{O}(n_c \cdot n_k)$.
The worst case for merging is when a community merges with all the others successively which could reach $\mathcal{O}(n_c \cdot n_k)$. This however can only happen at some specific scales when a mega community suddenly forms by absorbing the other communities. As this result (i.e. all nodes in one community) is not relevant to a community structure analysis it can be discarded. The merging is most cases consists in merging in linear time a set of pairs of communities. It is therefore expected to run with a complexity close to linear.

Overall the growth process is the part with the greatest complexity, running with a complexity between $\mathcal{O}(n_k \cdot d)$ and $\mathcal{O}(n_k \cdot d^2)$. Over all the communities $n_k \cdot n_c \ge n$ as  $n_k \cdot n_c = n$ when there is no overlap. Therefore $n_k \cdot n_c \cdot d \ge m$, with $m$ the number of edges. It gives an overall complexity in $\Omega(m)$ which represents the lowest bound when there is no overlap during computation.
In practice growing communities are expected to overlap and to potentially merge. Therefore the overlapping feature is used throughout the algorithm. Yet the overlapping is limited to a certain ratio and makes some nodes and edges processed more than once. It is thus expected to increase the constant factor only. We can therefore expect a linear complexity with respect to the number of edges .

Also throughout these operations described above some instructions operating on data structures (e.g. sets implemented as red-black trees) have a complexity of $log(n_k)$. As a result the overall complexity may be slightly super linear with an additional $log$ factor.

%%%%%%%%%%%%%%%%%%%%%%%%%%%%%%%%%%%%%%%%%%%%%%%%%%%%%%%%%%%%

\section{Experiments}

This section presents experiments that were performed to assess our algorithm. A dedicated implementation was coded in C++ (using C++11)\footnote{The code developed for this work is available for download from \protect\url{http://www.elemartelot.org}.}. All experiments were run under MacOS X 10.7.4 on a desktop computer iMac 3.06GHz Intel Core i3 with 4GB of RAM. The machine has 4 cores. Our implementation by default launches as many threads as there are cores. Therefore for these experiments it launches at most 4 threads for growth, checking and merging (see \aref{alg:lfk_growth_2}).

%In \cite{LeMartelot2013a} many experiments were run to assess the accuracy and the efficiency of the algorithms, including the one using the LFK criterion. 
%
%
%The aim of our method is to provide an efficient tool for the analysis of unknown potentially large networks. The algorithm must be accurate but also efficient to provide in a short amount of time some community sets that can potentially be further analysed using computational tools, visualisation methods, or any other relevant method. Therefore both accuracy and efficiency will be assessed.

In order to test the algorithm's performance and perform a comparative analysis of the criteria we used the benchmark from Lancichinetti et al. \cite{Lancichinetti2009a} that was designed to provide networks with communities at both micro and macro scales and encompassing properties found in real-life networks.

Regarding the scale parameters, we use a logarithmic sampling of the scale values within the interval of relevance to each criterion. The scale sampling is given by
\[Values = V_{min} + (A-V_{min}) \cdot \frac{1-log([1:1:X])}{log(X)}\]
where X is the number of values we want in the sample, $[1:1:X]$ the vector of values between $1$ and $X$ incremented by $1$ between each value. The formula returns a vector of $X$ sample values within the interval $[V_{min},A]$ with values around $V_{min}$ close to each other and then progressively spreading out towards $A$.

The information change between community sets is measured using the normalised mutual information (NMI) for overlapping communities from \cite{Lancichinetti2009b} which is an alternative definition to the one from \cite{Fred2003}.
To analyse how much change there is between successive community sets we measure the NMI averaged over $p$ successive scales. We use $p=3$ and $p=5$ in our experiments. A short range reveals a potentially short consistency between community sets while a longer range reveals longer consistencies. The longer the consistency the more robust to scale variation a community set is, and the more confidence we can have in the relevance of the set.

\subsection{Accuracy}

\begin{figure*}[htb]
	\centering
	\subfigure[New LFK with multi-threading]{
		\includegraphics[width=0.48\textwidth]{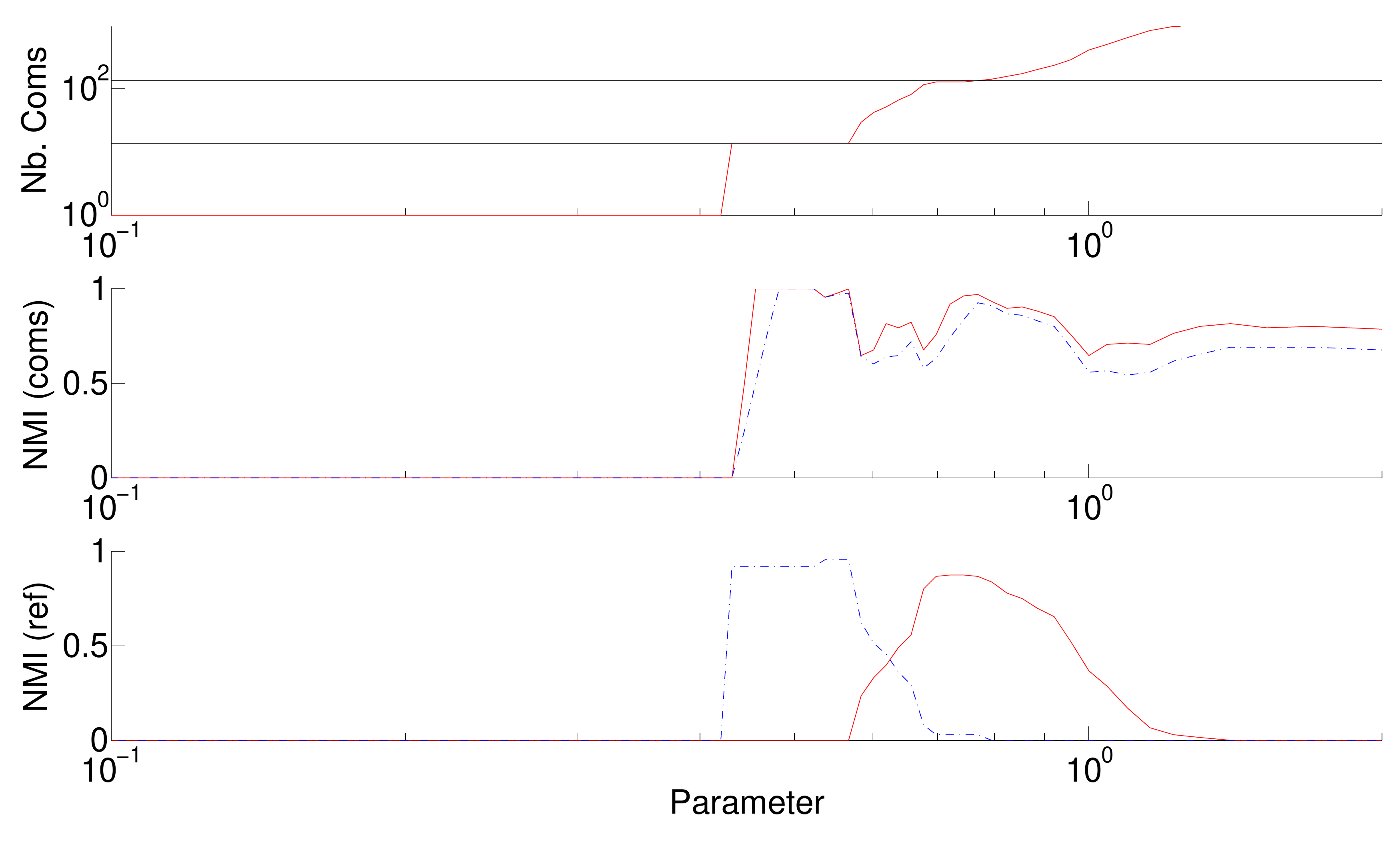}
		\label{fig:n10p4_05_2_LFK2}
	}
	\subfigure[LFK from previous work \cite{LeMartelot2013a}]{
		\includegraphics[width=0.48\textwidth]{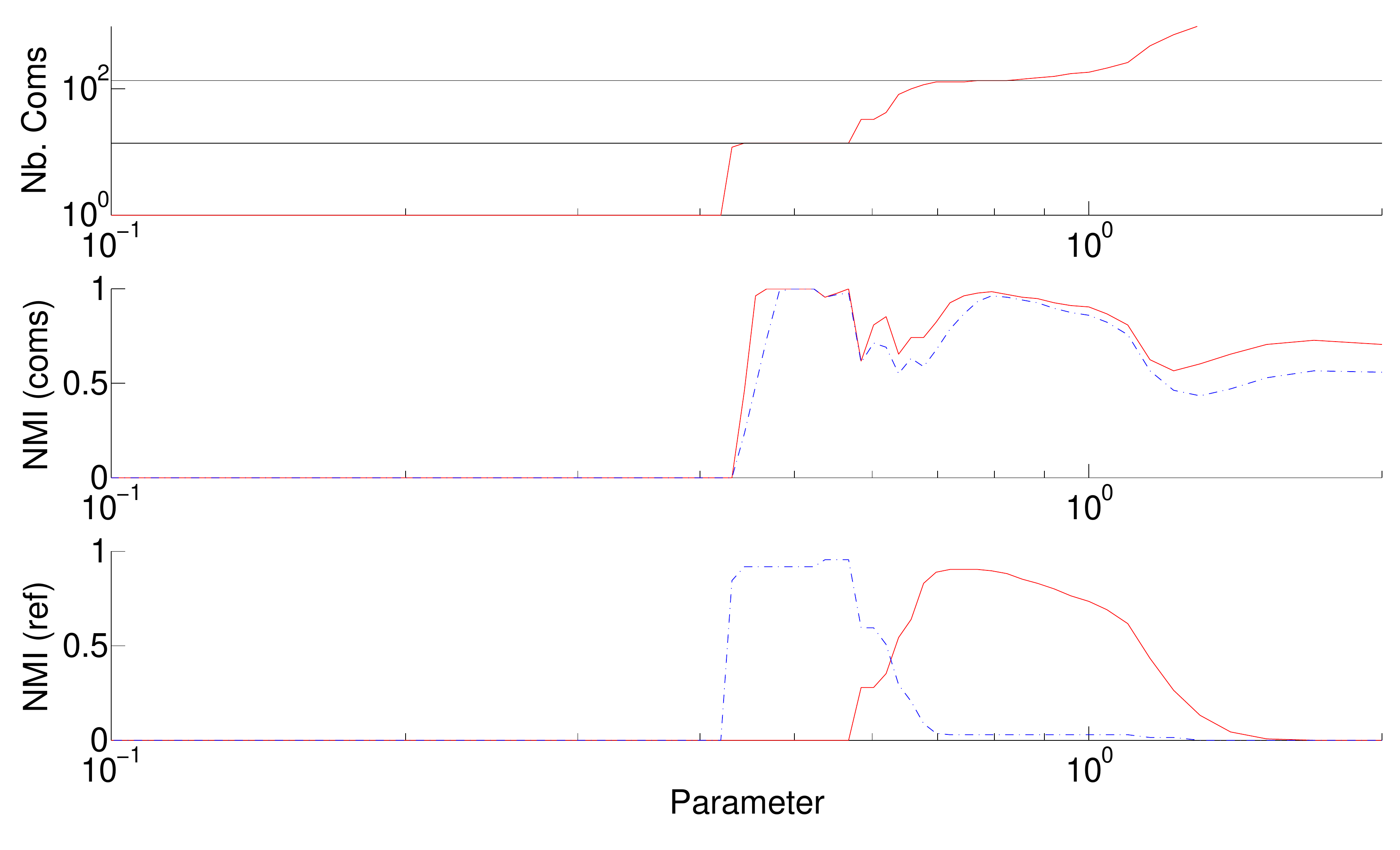}
		\label{fig:n10p4_05_2_LFK}
	}
	\subfigure[New LFK with only 1 thread]{
		\includegraphics[width=0.48\textwidth]{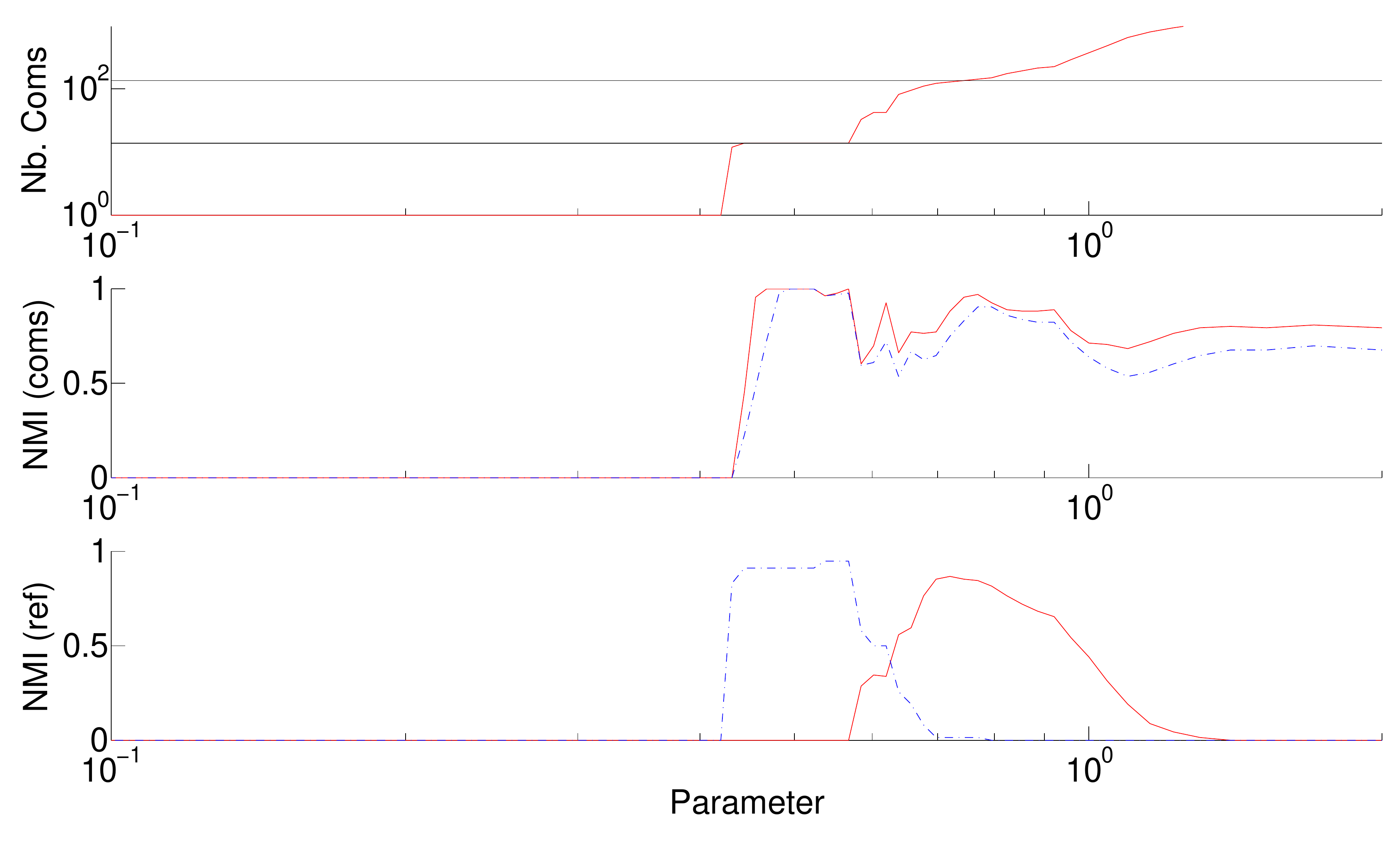}
		\label{fig:n10p4_05_2_LFK2t1}
	}
	\subfigure[New LFK with multi-threading and $2^{nd}$ seed rule]{
		\includegraphics[width=0.48\textwidth]{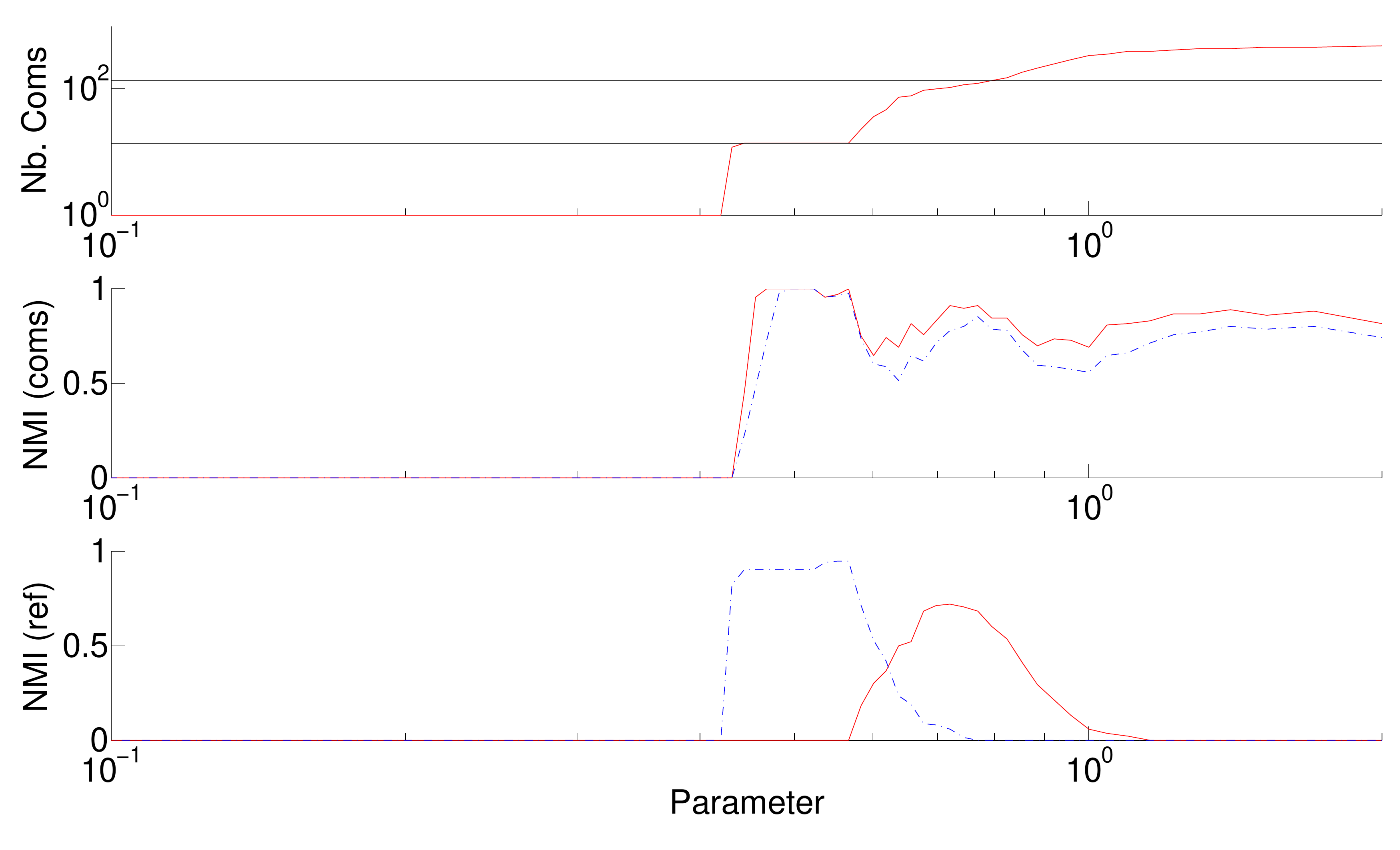}
		\label{fig:n10p4_05_2_LFK2s2}
	}
	\caption{Result analysis along the scale parameter using a generated network with $n=10^4$ nodes, $m \approx 10^5$, $\mu_1 = 0.05$ and $\mu_2 = 0.2$. The top plot indicates the number of communities uncovered with the two intended community set's size shown in black straight lines. The second plot shows the averaged NMI between successive uncovered sets of communities: 3 in (red) full and 5 in (blue) dashed. The third plot shows the NMI with the two intended partitions: in (red) full the micro communities and in (blue) dashed the macro communities. The results are presented for \subref{fig:n10p4_05_2_LFK2} the new algorithm, \subref{fig:n10p4_05_2_LFK} the initial algorithm from \cite{LeMartelot2013a}, \subref{fig:n10p4_05_2_LFK2t1} the new algorithm using only one thread, and \subref{fig:n10p4_05_2_LFK2s2} the new algorithm using the second seed rule.}
	\label{fig:n10p4_05_2_res}
\end{figure*}

In this sets of experiments we compare the accuracy of the initial LFK algorithm from \cite{LeMartelot2013a} with our new algorithm designed for multi-threading. We use three setups for our new algorithm. The first setup is the default setup using multi-threading (4 threads on our machine). The second one uses only one thread in order to assess the algorithm in a non multi-threaded environment. The third setup uses multi-threading but initialises the seeds with the second rule (not allowing neighbours of neighbours of seeds to be seeds). \fref{fig:n10p4_05_2_res} shows the results of the multi-scale analysis on a generated network with $10^4$ nodes, about $10^5$ edges, $\mu_1 = 0.05$ and $\mu_2 = 0.2$. Therefore 5\% and 20\% of the edges belonging respectfully to the macro and micro communities point outside their communities.

Overall the micro and macro communities are well detected by all setups. The new algorithm, whether using multi-threading or just one thread, detects well the micro and macro communities. However we can observe on \fref{fig:n10p4_05_2_LFK2s2} that the setup using the second seed rule detects the micro communities with less accuracy. This is visible on the NMI with the reference communities plot where the NMI value peak (around scale 0.75) is lower for the micro-communities than the same peak for the other setups. Similarly the NMI across successive communities peaks at a lower height than with the other setups. Hence while there is a detection of micro-communities between scales $0.7$ and $0.8$, the second seed rule version is less confident in its detection (NMI across community sets) and indeed less accurate (NMI with reference communities) than the other setups.
The detection of macro communities is however as accurate as with the other setups. As the selection of seeds is coarser, the analysis at a micro scale may then be coarser, hence a less accurate detection in micro-communities and a similar accuracy for macro communities.

The same experiment is repeated setting $\mu_1 = 0.2$ and $\mu_2 = 0.4$. Therefore there is significantly more noise in the communities: 20\% and 40\% of the edges belonging respectfully to the macro and micro communities point outside their communities.

\begin{figure*}[htb]
	\centering
	\subfigure[New LFK with multi-threading]{
		\includegraphics[width=0.48\textwidth]{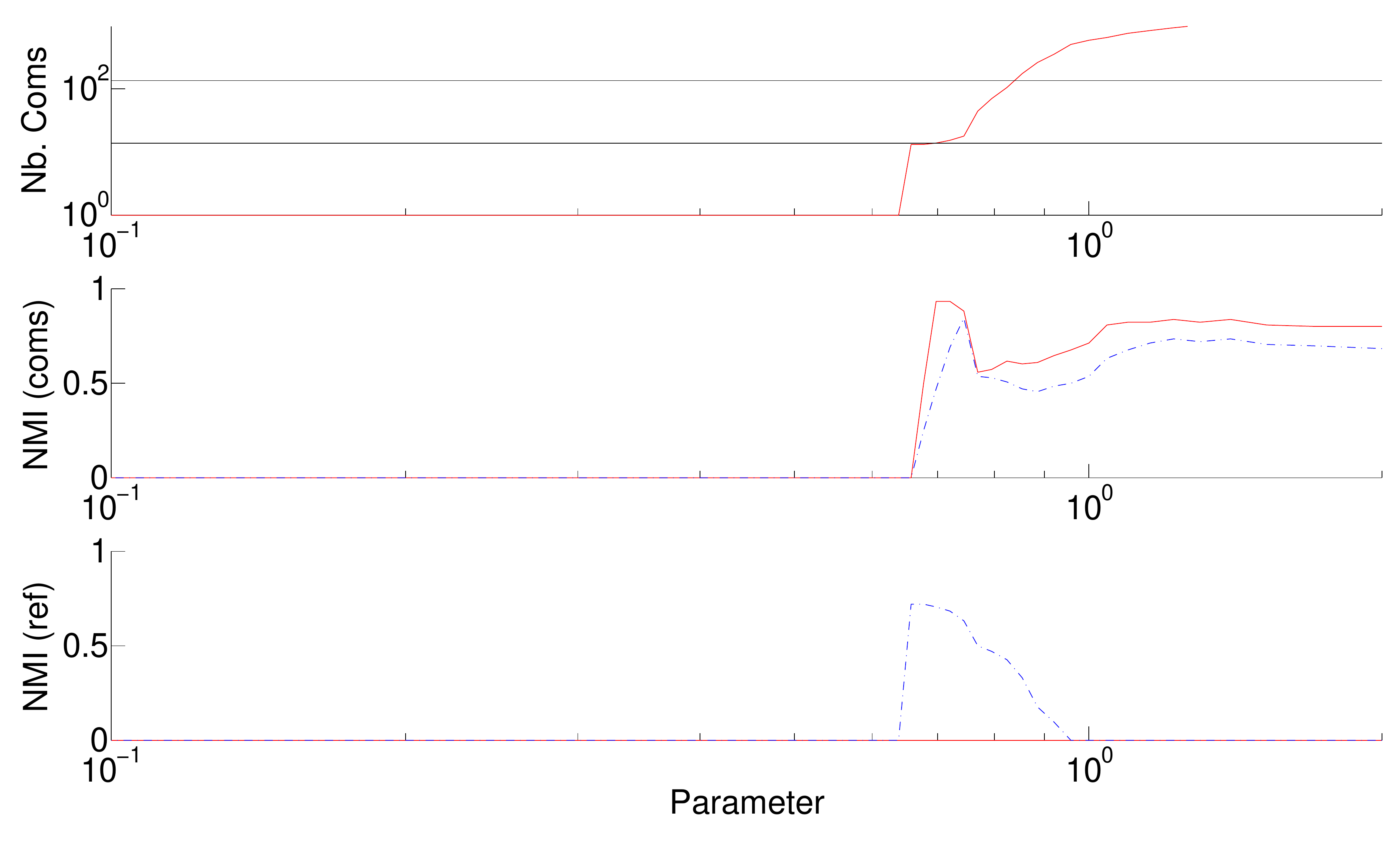}
		\label{fig:n10p4_2_4_LFK2}
	}
	\subfigure[LFK from previous work \cite{LeMartelot2013a}]{
		\includegraphics[width=0.48\textwidth]{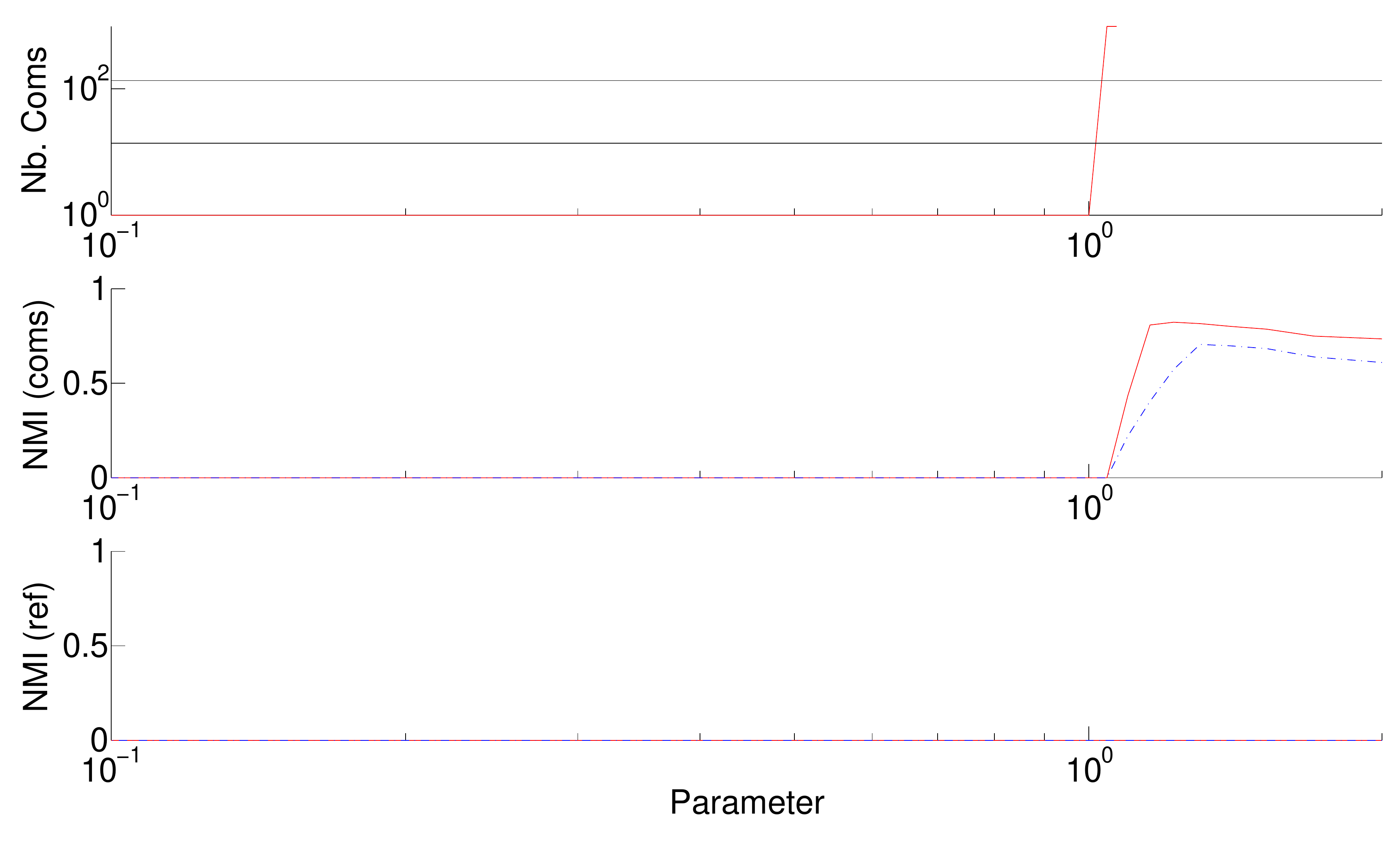}
		\label{fig:n10p4_2_4_LFK}
	}
	\subfigure[New LFK with only 1 thread]{
		\includegraphics[width=0.48\textwidth]{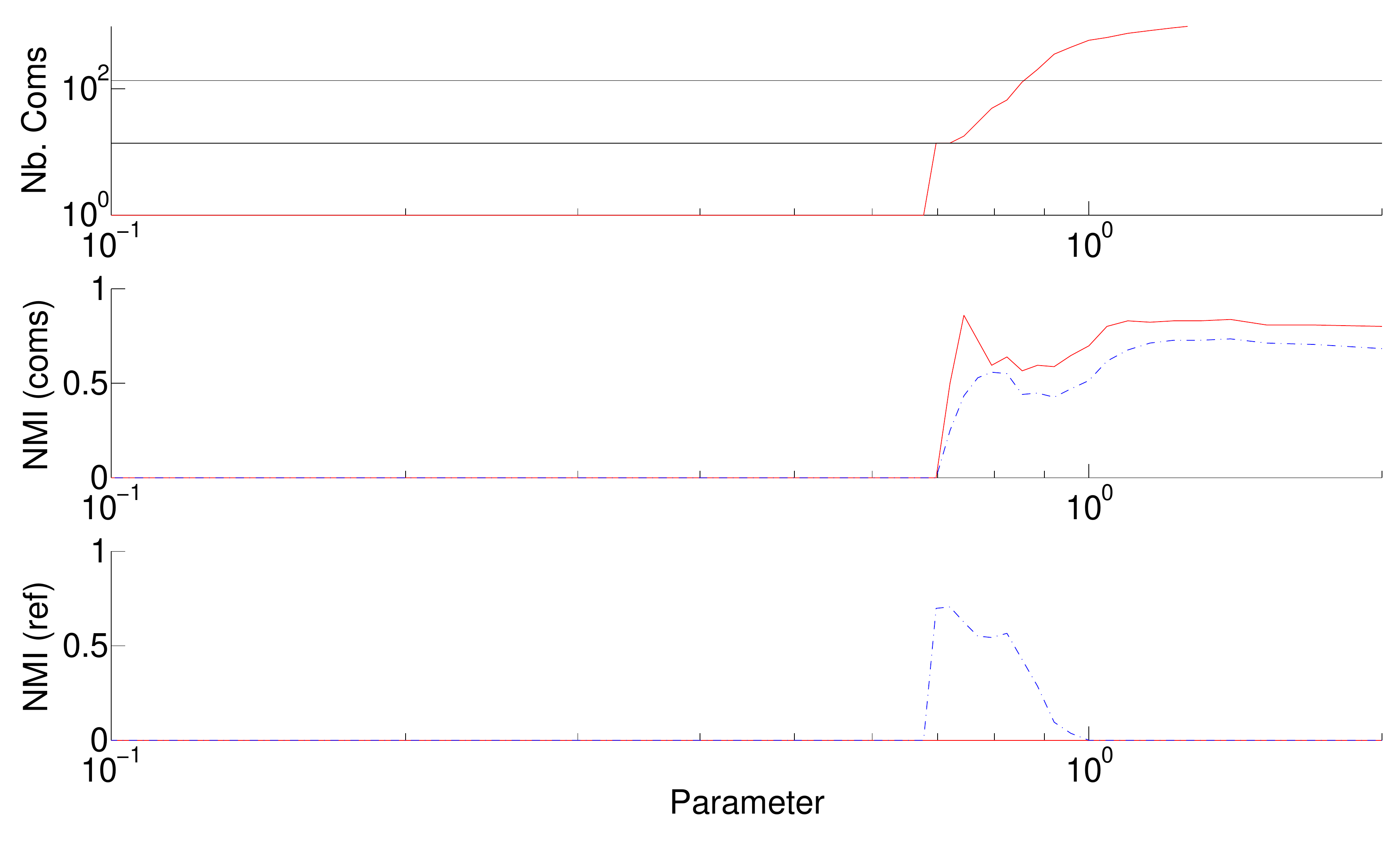}
		\label{fig:n10p4_2_4_LFK2t1}
	}
	\subfigure[New LFK with multi-threading and $2^{nd}$ seed rule]{
		\includegraphics[width=0.48\textwidth]{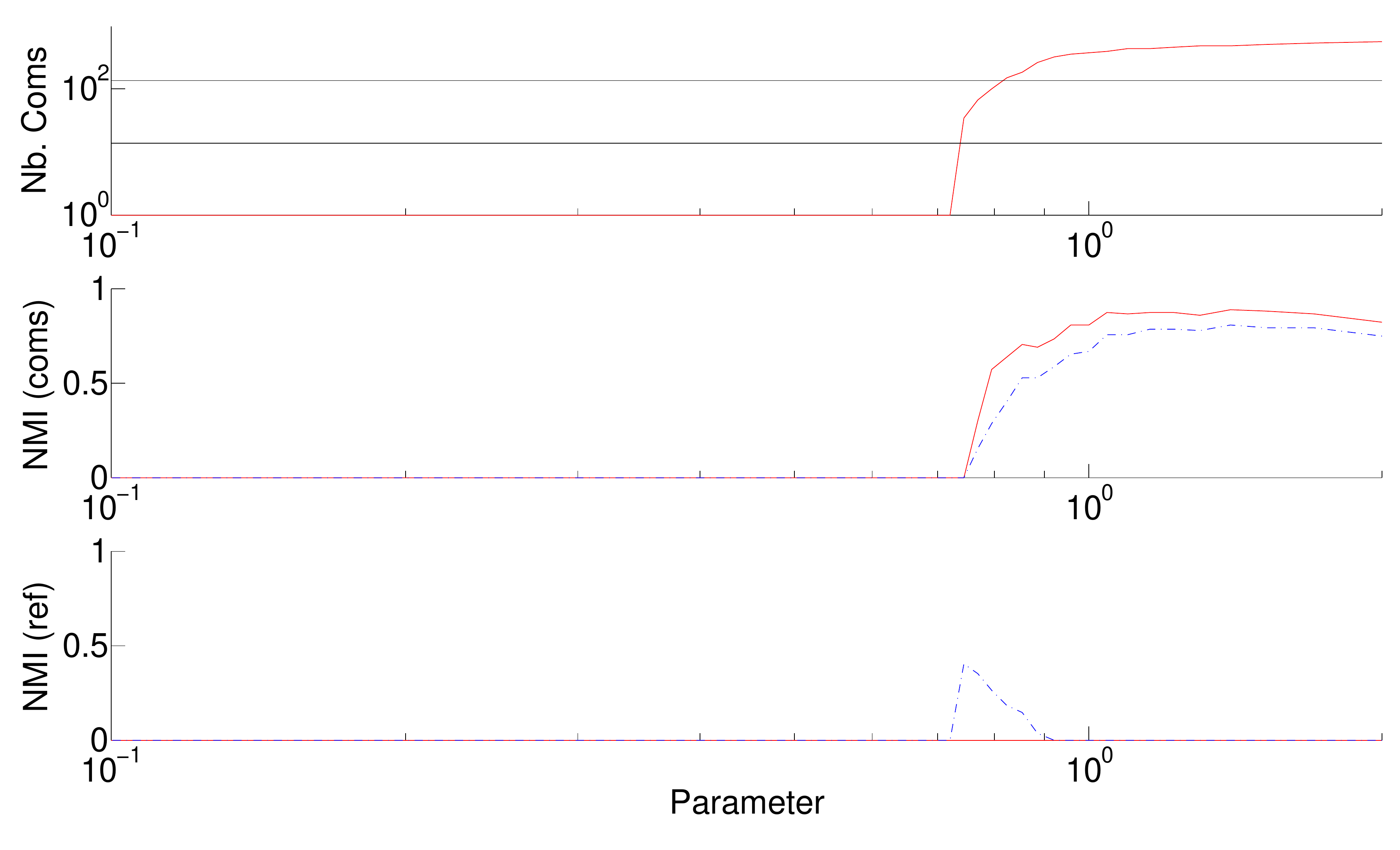}
		\label{fig:n10p4_2_4_LFK2s2}
	}
	\caption{Result analysis along the scale parameter using a generated network with $n=10^4$ nodes, $m \approx 10^5$, $\mu_1 = 0.2$ and $\mu_2 = 0.4$. The top plot indicates the number of communities uncovered with the two intended community set's size shown in black straight lines. The second plot shows the averaged NMI between successive uncovered sets of communities: 3 in (red) full and 5 in (blue) dashed. The third plot shows the NMI with the two intended partitions: in (red) full the micro communities and in (blue) dashed the macro communities. The results are presented for \subref{fig:n10p4_05_2_LFK2} the new algorithm, \subref{fig:n10p4_05_2_LFK} the initial algorithm from \cite{LeMartelot2013a}, \subref{fig:n10p4_05_2_LFK2t1} the new algorithm using only one thread, and \subref{fig:n10p4_05_2_LFK2s2} the new algorithm using the second seed rule.}
	\label{fig:n10p4_2_4_res}
\end{figure*}

On \fref{fig:n10p4_2_4_res} we first observe that the initial algorithm from \cite{LeMartelot2013a} does not detect any community set. Indeed the number of communities suddenly drops from several hundreds to only one. This suggests that all communities suddenly merged into one at a given scale parameter value. This algorithm indeed checks all possible combinations of communities to merge at each merging step. Even though this can provide a better accuracy in ensuring that all communities overlapping significantly are properly merged, it can also lead to the premature emergence of a mega community, thus missing relevant divisions.
The new algorithm presented in this work avoids this drawback by only allowing communities that have grown to be checked for merging. This means that if a community has not grown, it cannot join another community. Another community that was grown can join it though.

The next observation is that the new algorithm clearly detects the macro communities with the original seed rule. When using the second seed rule these communities are detected but not very clearly. Therefore a coarse initial selection of seeds can also reduces the accuracy of the analysis on macro communities.

Finally it is noteworthy that the micro-communities are not detected by any method. Experiments show that the technique based on this criterion is not very resistant to noise. This is consistent with the results from \cite{LeMartelot2013a} that showed that global criteria approaches cope better with noise than local approaches (within the scope of the criteria under study).

To investigate further the accuracy of the algorithms based on the amount of noise introduced in the communities, the algorithm has been run on various networks varying the values of $\mu_1$ and $\mu_2$. These results are summarised below in \tref{tab:acc_exp}.

\begin{table*}[hbt]
	\caption{Scale parameter range where the macro and then micro communities were spotted. Clearly identified ranges use the interval notation $[ ]$, values of interest with no clear stable range but a clear NMI peak (weak detection) are given using the notation $()$ and the empty set denotes no detection of the community scale. The first network's results are shown in \fref{fig:n10p4_05_2_res} and the third network's results are shown in \fref{fig:n10p4_2_4_res}.}
	\label{tab:acc_exp}
	\begin{center}
	\mbox{Networks with $n=10^4$ and $m \approx 10^5$} \vspace{2mm}
	\footnotesize{
	\begin{tabular}{|l|l|l|l|l|} \hline
		Criteria		& $\mu_1=0.05$, $\mu_2=0.2$		& $\mu_1=0.05$, $\mu_2=0.3$		& $\mu_1=0.2$, $\mu_2=0.4$	& $\mu_1=0.3$, $\mu_2=0.4$\\ \hline
		LFK2		& [0.44,0.58] [0.7,0.8]					& [0.39,0.80] (0.85,0.95)				& [0.75] $\emptyset$					& $\emptyset$ $\emptyset$\\
		LFK2 1th	& [0.44,0.58] [0.7,0.8]					& [0.44,0.80] (0.9,0.95)					& [0.7,0.71] $\emptyset$			& $\emptyset$ $\emptyset$\\
		LFK2 sr2	& [0.44,0.58] [0.7,0.8]					& [0.39,0.70] (0.9)							& (0.72) $\emptyset$					& $\emptyset$ $\emptyset$\\
		LFK			& [0.44,0.58] [0.7,0.8]					& [0.44,0.76] (0.9,1.0)					& $\emptyset$ $\emptyset$		& $\emptyset$ $\emptyset$\\ \hline
	\end{tabular}
	}
	\end{center}
\end{table*}

\subsection{Speed Performance and Memory Usage}

\begin{figure*}[htb]
	\centering
	\subfigure[Speed performance / Network size]{
		\includegraphics[width=0.48\textwidth]{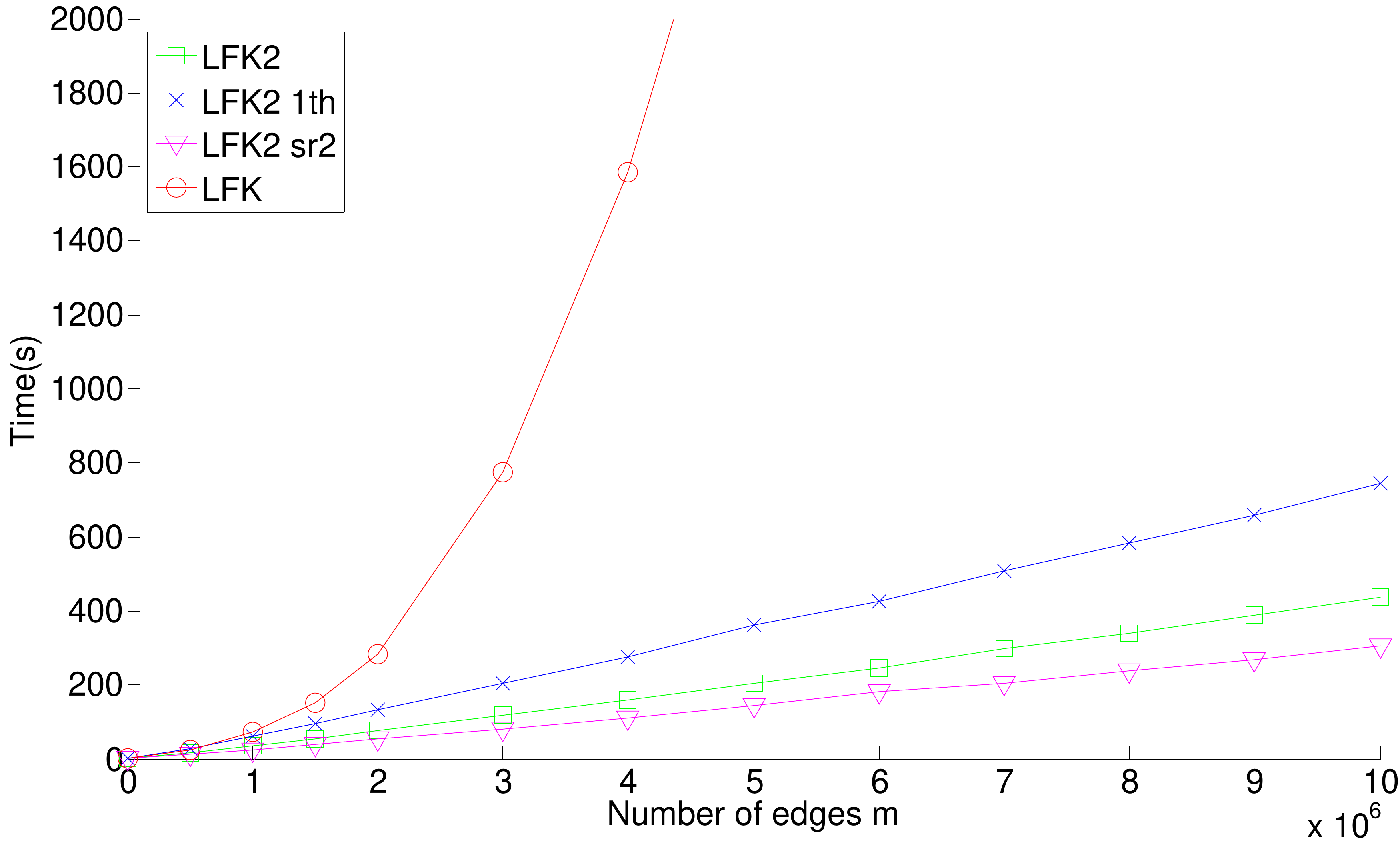}
		\label{fig:speed_size_exp}
	}
	\subfigure[Memory usage / Network size]{
		\includegraphics[width=0.48\textwidth]{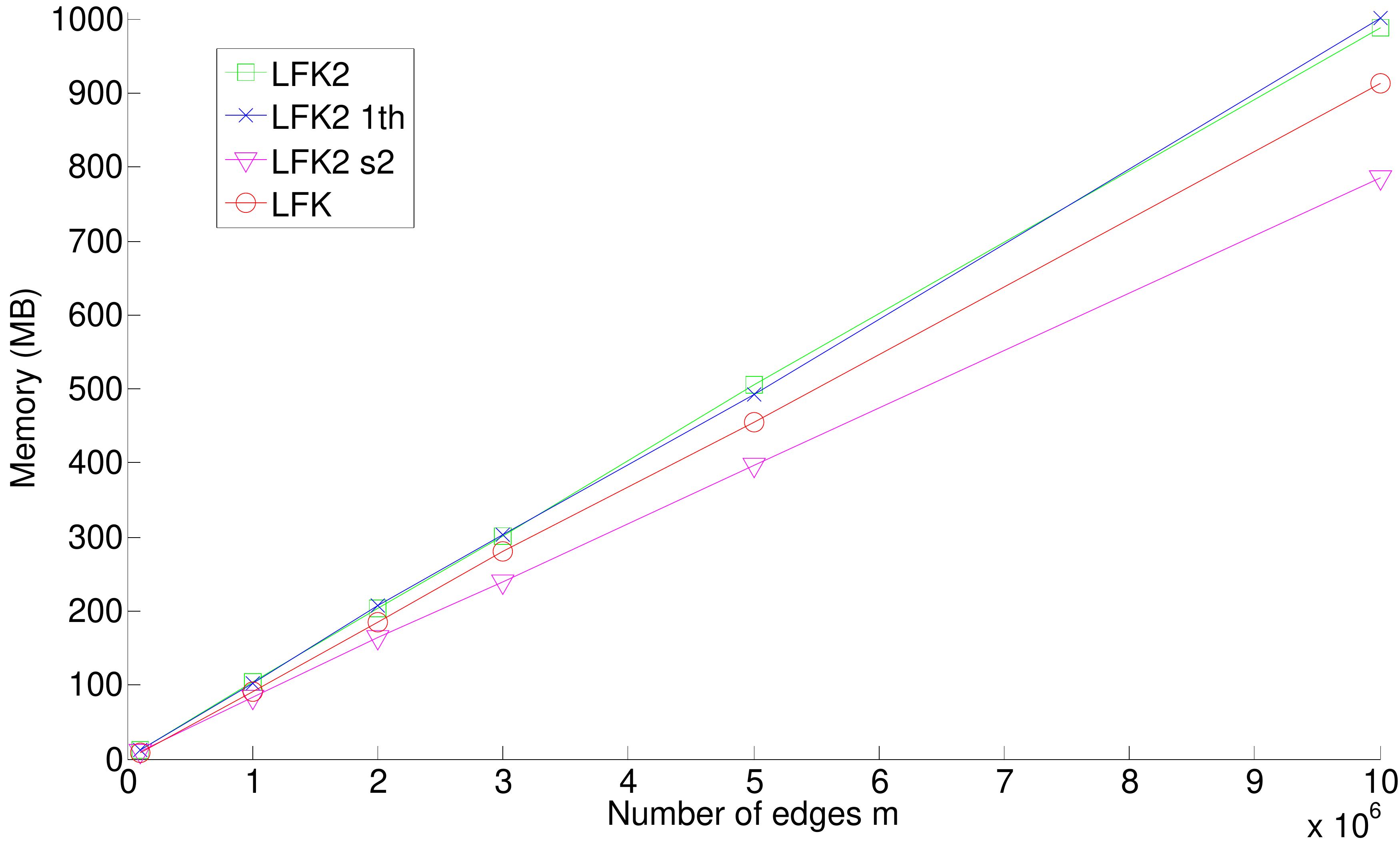}
		\label{fig:memory_size_exp}
	}
	\caption{Speed performance \subref{fig:speed_size_exp} and memory usage \subref{fig:memory_size_exp} for the original and the new algorithm given the network's size in edges $m \approx 10n$ up to networks with $m=10^7$.}
	\label{fig:size_exp}
\end{figure*}

To assess the scalability of our algorithm we used networks with between $10^4$ and $10^7$ edges generated with $\mu_1=0.1$ and $\mu_2=0.2$. We also set $V_{min} = 0.5$ and $A=1$. Indeed, if we consider \fref{fig:n10p4_05_2_res} and the results from \tref{tab:acc_exp}, all the relevant detection is done by $\alpha = 0.5$ and lower scale values have either the same set of communities or one community. Somewhere between $0.4$ and $0.5$, all communities merge into one. Such a set of operations may become sequential with the creation of a mega-community absorbing others and is of limited interest to assess parallel computation. Therefore we run the speed experiments on scales where a significant amount of communities co-exist and where parallel computation can take place. The results are presented in \fref{fig:speed_size_exp}.

We can observe that our new algorithm runs with a linear complexity, as expected from the complexity study, while the initial algorithm from \cite{LeMartelot2013a} runs with a polynomial complexity. Therefore our new algorithm can process a network of $10^7$ edges over 100 scales in about 7 minutes. Using only one thread the same network can be processed in about 12 minutes. The algorithm can therefore run very efficiently on mono-processor machines. Finally the version of our new algorithm using the second seed rule runs even faster and can process the same network in about 5 minutes. This result is comparable to and even faster than the fastest result obtained in \cite{LeMartelot2013a} with the global algorithm which does not allow for overlapping communities.
Our new algorithm thus brought the local criterion based algorithm with overlapping communities to the same complexity and efficiency as the global criterion algorithms.

Memory usage is also measured and given in \fref{fig:memory_size_exp}. The larger the networks, the more memory needed to store them and any related data structure the algorithm requires. We expect and can observe a usage growing linearly with the network size.
The new algorithm requires a bit of additional memory compared to the one from \cite{LeMartelot2013a} due to the threads data structures.
We also observe that the version of the algorithm using the second seed rule consistently uses less memory than the version using the regular seed rule. As the number of communities to grow is reduced, the amount of memory needed is lessened.

%%%%%%%%%%%%%%%%%%%%%%%%%%%%%%%%%%%%%%%%%%%%%%%%%%%%%%%%%%%%

%%%%%%%%%%%%%%%%%%%%%%%%%%%%%%%%%%%%%%%%%%%%%%%%%%%%%%%%%%%%

\section{Conclusion}

In this paper we presented an algorithm for the fast detection of communities across scales using a local criterion. This work is based on previous work \cite{LeMartelot2013a} which introduced a method for the fast multi-scale detection of communities. The method was implemented into two algorithms: one designed for global criteria and one designed for local criteria. However the local criteria implementation was significantly less efficient than the global criteria implementation. Indeed the local criteria based implementations allow communities to overlap which increases the complexity of the task.
In this work we addressed this issue by introducing a new algorithm based on the method from \cite{LeMartelot2013a} and designed for multi-threading. Complexity analysis showed that the new algorithm is expected to run with linear complexity, as opposed to the initial algorithm that exhibits polynomial complexity. Experiments corroborated this theoretical result and demonstrated the improved efficiency but also accuracy of the new algorithm over the algorithm from \cite{LeMartelot2013a}. Experiments also showed that the algorithm remains very efficient without using parallelism (i.e. running a single thread). The addition of threads hence lowers the overall running time.
It is expected that a largely parallel architecture can enable significant speed-ups. Also two initialisation rules were suggested for the creation of the initial communities. The first rule leads to the best accuracy while the second rule may sacrifice a bit of accuracy, particularly when communities have many edges pointing outside (e.g. a lot of noise), for an improved efficiency. Using the second seed rule our algorithm runs faster than the fastest global criterion algorithm from \cite{LeMartelot2013a}, making it the fastest implementation of the fast multi-scale community detection method.

%%%%%%%%%%%%%%%%%%%%%%%%%%%%%%%%%%%%%%%%%%%%%%%%%%%%%%%%%%%%

% Acknowledgements
\section*{Acknowledgements}
This work was conducted as part of the \emph{Making Sense}\footnote{Making sense project: \url{http://www.making-sense.org}.} project financially supported by EPSRC (project number EP/H023135/1).

% Bibliography style and file
\bibliographystyle{plainnat-nourl}
\bibliography{../../Bibliography/Bibliography}

\begin{thebibliography}{18}
\expandafter\ifx\csname natexlab\endcsname\relax\def\natexlab#1{#1}\fi
\expandafter\ifx\csname url\endcsname\relax
  \def\url#1{{\tt #1}}\fi

\bibitem[Arenas et~al.(2008)Arenas, Fernandez, and Gomez]{Arenas2008}
Alex Arenas, Alberto Fernandez, and Sergio Gomez.
\newblock Analysis of the structure of complex networks at different resolution
  levels.
\newblock {\em New Journal of Physics}, 10:\penalty0 053039, Jan 2008.

\bibitem[Courtois et~al.(1971)Courtois, Heymans, and Parnas]{Courtois1971}
Pierre~J. Courtois, F.~Heymans, and David~L. Parnas.
\newblock Concurrent control with ``readers'' and ``writers''.
\newblock {\em Communications of the ACM}, 14\penalty0 (10):\penalty0 667--668,
  October 1971.
\newblock ISSN 0001-0782.

\bibitem[Fortunato(2010)]{Fortunato2010}
Santo Fortunato.
\newblock Community detection in graphs.
\newblock {\em Physics Reports}, 486\penalty0 (3-5):\penalty0 75--174, 2010.
\newblock ISSN 0370-1573.

\bibitem[Fred and Jain(2003)]{Fred2003}
Ana L.~N. Fred and Anil~K. Jain.
\newblock Robust data clustering.
\newblock {\em IEEE Computer Society Conference on Computer Vision and Pattern
  Recognition}, 2:\penalty0 128--133, 2003.
\newblock ISSN 1063-6919.

\bibitem[Huang et~al.(2011)Huang, Sun, Liu, Song, and Weninger]{Huang2011}
Jianbin Huang, Heli Sun, Yaguang Liu, Qinbao Song, and Tim Weninger.
\newblock Towards online multiresolution community detection in large-scale
  networks.
\newblock {\em PLoS ONE}, 6\penalty0 (8):\penalty0 e23829, 08 2011.

\bibitem[Lancichinetti and Fortunato(2009)]{Lancichinetti2009a}
Andrea Lancichinetti and Santo Fortunato.
\newblock Benchmarks for testing community detection algorithms on directed and
  weighted graphs with overlapping communities.
\newblock {\em Physical Review E}, 80\penalty0 (1):\penalty0 016118, 2009.

\bibitem[Lancichinetti et~al.(2009)Lancichinetti, Fortunato, and
  Kert\'esz]{Lancichinetti2009b}
Andrea Lancichinetti, Santo Fortunato, and J\'anos Kert\'esz.
\newblock Detecting the overlapping and hierarchical community structure in
  complex networks.
\newblock {\em New Journal of Physics}, 11\penalty0 (3):\penalty0 033015, 2009.

\bibitem[Le~Martelot and Hankin(2013{\natexlab{a}})]{LeMartelot2013a}
Erwan Le~Martelot and Chris Hankin.
\newblock Fast multi-scale detection of relevant communities in large scale
  networks.
\newblock {\em The Computer Journal}, 2013{\natexlab{a}}.

\bibitem[Le~Martelot and Hankin(2013{\natexlab{b}})]{LeMartelot2012}
Erwan Le~Martelot and Chris Hankin.
\newblock Multi-scale community detection using stability optimisation.
\newblock {\em The International Journal of Web Based Communities (IJWBC)
  Special Issue on Community Structure in Complex Networks}, 9\penalty0 (3),
  2013{\natexlab{b}}.

\bibitem[Leskovec et~al.(2008)Leskovec, Lang, Dasgupta, and
  Mahoney]{Leskovec2008}
Jure Leskovec, Kevin~J. Lang, Anirban Dasgupta, and Michael~W. Mahoney.
\newblock Statistical properties of community structure in large social and
  information networks.
\newblock In {\em Proceeding of the 17th international conference on World Wide
  Web}, WWW '08, pages 695--704, New York, NY, USA, 2008. ACM.
\newblock ISBN 978-1-60558-085-2.

\bibitem[Leskovec et~al.(2010)Leskovec, Lang, and Mahoney]{Leskovec2010}
Jure Leskovec, Kevin~J. Lang, and Michael Mahoney.
\newblock Empirical comparison of algorithms for network community detection.
\newblock In {\em Proceedings of the 19th international conference on World
  wide web}, WWW '10, pages 631--640, New York, NY, USA, 2010. ACM.
\newblock ISBN 978-1-60558-799-8.

\bibitem[Newman(2010)]{Newman2010}
Mark E.~J. Newman.
\newblock {\em Networks, an Introduction}.
\newblock Oxford University Press, 1st edition, 2010.

\bibitem[Raghavan et~al.(2007)Raghavan, Albert, and Kumara]{Raghavan2007}
Usha~Nandini Raghavan, R\'eka Albert, and Soundar Kumara.
\newblock Near linear time algorithm to detect community structures in
  large-scale networks.
\newblock {\em Physical Review E}, 76:\penalty0 036106, Sep 2007.

\bibitem[Reichardt and Bornholdt(2006)]{Reichardt2006}
J.~Reichardt and S.~Bornholdt.
\newblock Statistical mechanics of community detection.
\newblock {\em Physical Review E}, 74\penalty0 (1 Pt 2):\penalty0 016110, July
  2006.
\newblock ISSN 1539-3755.

\bibitem[Riedy et~al.(2012)Riedy, Meyerhenke, Ediger, and Bader]{Riedy2012b}
E.~Jason Riedy, Henning Meyerhenke, David Ediger, and David~A. Bader.
\newblock Parallel community detection for massive graphs.
\newblock In {\em 10th DIMACS Implementation Challenge - Graph Partitioning and
  Graph Clustering}. Atlanta, Georgia, February 2012.

\bibitem[Ronhovde and Nussinov(2010)]{Ronhovde2010}
Peter Ronhovde and Zohar Nussinov.
\newblock Local resolution-limit-free {P}otts model for community detection.
\newblock {\em Physical Review E}, 81\penalty0 (4):\penalty0 046114, April
  2010.

\bibitem[Simon(1962)]{Simon1962}
Herbert~A. Simon.
\newblock The architecture of complexity.
\newblock {\em Proceedings of the American Philosophical Society}, 106\penalty0
  (6):\penalty0 467--482, December 1962.

\bibitem[Soman and Narang(2011)]{Soman2011}
J.~Soman and A.~Narang.
\newblock Fast community detection algorithm with gpus and multicore
  architectures.
\newblock In {\em Parallel Distributed Processing Symposium (IPDPS), 2011 IEEE
  International}, pages 568 --579, may 2011.

\end{thebibliography}

\end{document}